\documentclass[runningheads]{llncs}
\usepackage[T1]{fontenc}
\usepackage{graphicx}
\usepackage{cite}
\usepackage{amsmath,amssymb,amsfonts}
\usepackage{algpseudocode}
\usepackage{algorithmicx}
\usepackage{textcomp}
\usepackage{xcolor}
\usepackage{subcaption}
\usepackage{float}
\newcommand{\highlight}[1]{\textbf{#1}}

\usepackage{hyperref}
\usepackage{CJKutf8}
\usepackage{multirow}
\usepackage{graphicx} 
\usepackage{pgfplots}
\usepackage{tikz}
\usepackage{booktabs}
\usepackage{flushend}
\usepackage{amsmath}
\usepackage{algorithm}
\usepackage{algpseudocode}
\usepackage{amssymb}
\usepackage{adjustbox}
\usepackage{array}
\usepackage{pifont}     

\usepackage{tabularx}
\usepackage{booktabs}
\usepackage{hyperref}
\usepackage{flushend}
\titlerunning{MAS4POI}
\title{MAS4POI: a Multi-Agents Collaboration System for Next POI Recommendation}

\author{
Yuqian Wu\textsuperscript{1}, Yuhong Peng\textsuperscript{1}, Jiapeng Yu\textsuperscript{2}, and Raymond S. T. Lee\textsuperscript{1,}\textsuperscript{\textdagger}, Member, IEEE
}
\authorrunning{Wu et al.}
\institute{
\textsuperscript{1}Beijing Normal University-Hong Kong Baptist University United International College, Zhuhai 519000, China\\
\textsuperscript{2}University of Warwick, Coventry CV4 7AL, United Kingdom\\
\email{r130034042@mail.uic.edu.cn, r130034029@mail.uic.edu.cn, Jiapeng.Yu@warwick.ac.uk}\\
\textsuperscript{\textdagger}\textit{Corresponding author: raymondshtlee@uic.edu.cn}
}

\date{}

\begin{document} 

\maketitle 

\vspace{-1.0em}
\begin{abstract} 
LLM-based Multi-Agent Systems have potential benefits of complex decision-making  tasks management across various  domains but their applications in the next Point-of-Interest (POI) recommendation remain underexplored.
This paper proposes a novel MAS4POI system designed to enhance next POI recommendations through multi-agent interactions. MAS4POI supports Large Language Models (LLMs) specializing in distinct agents such as DataAgent, Manager, Analyst, and Navigator with each contributes to a collaborative process of generating the next POI recommendations.
The system is examined by integrating six distinct LLMs and evaluated by two real-world datasets for recommendation accuracy improvement in real-world scenarios.
Our code is available at \url{https://github.com/yuqian2003/MAS4POI}.

\vspace{1.0em}
\end{abstract}

\section{Introduction}
\label{sec:intro}
\begin{figure}[h!]
    \centering
    \includegraphics[width=0.8\textwidth]{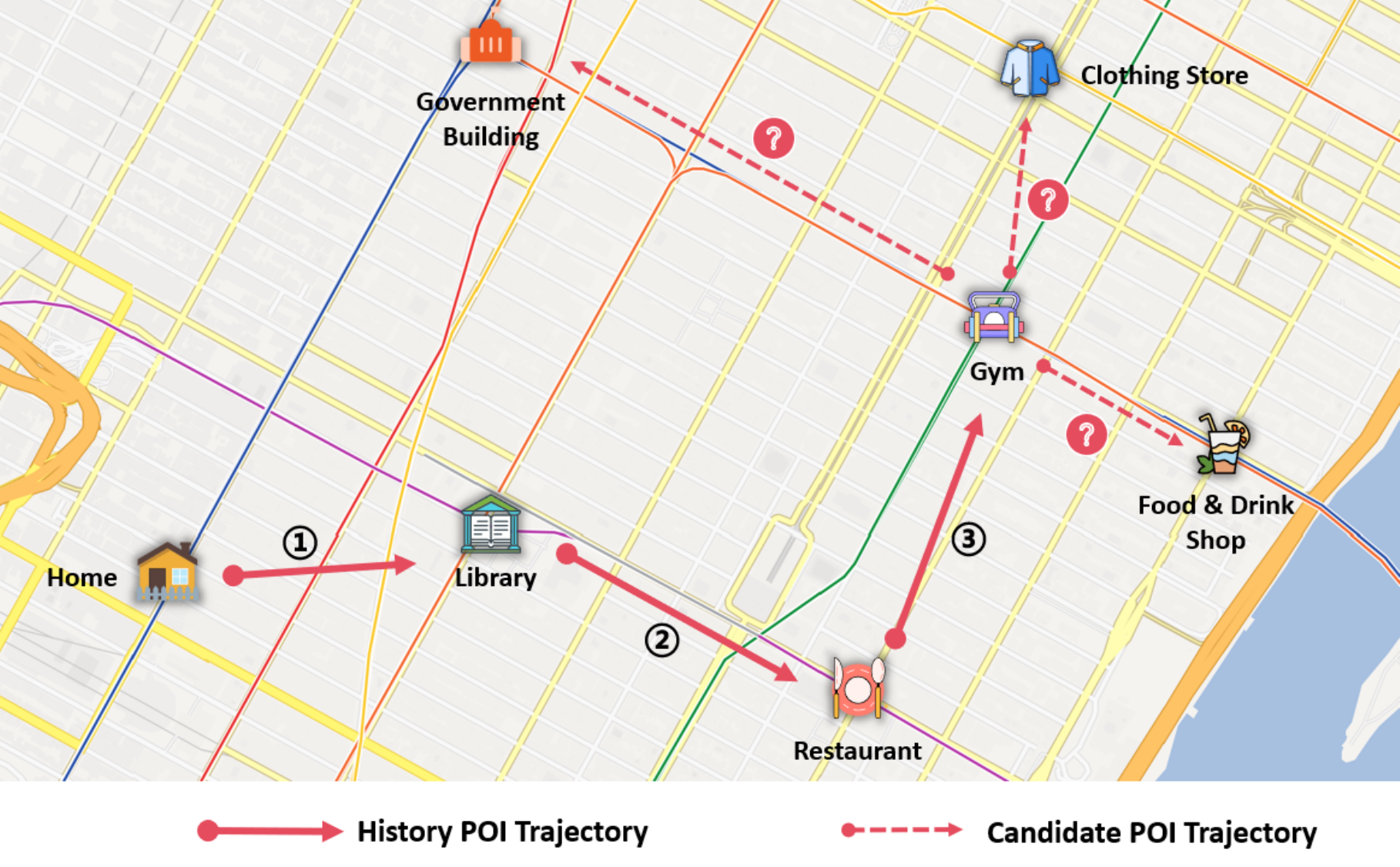}
    \caption{Illustration of a user's historical trajectory (solid line) and the candidate POIs for the next visit (dashed line).}
    \label{fig:map}
    \vspace{-2.5em}
\end{figure}

Location-Based Social Networks (LBSNs) like Foursquare and Facebook Places platforms have induced smartphone technology advancement by providing   users social opportunities to share check-in records, images, and Points of Interest (POIs) reviews.  
They have generated a wealth of data and new avenues for academia and industrial research on POI recommendation systems, a subfield of recommender systems.
The next POI Recommendation are designs to predict a user’s probable location supported by historical data such as geographical contexts, temporal patterns and personal preferences. 
An illustration of a user’s historical trajectory is shown in Fig.~\ref{fig:map}.

LLM-based multi-agent systems (MAS) specialize LLMs into distinct agents with each contributes to functionalities, and tailor the interactions among these agents for complex real-world tasks management [5]. LLM-based MAS collaborate with planning, discussion, decision-making, mirroring human group endeavour in problem-solving are effective in  software development~\cite{2024soft}, multi-robot systems~\cite{2023robot}, social simulation~\cite{2024self}, gaming~\cite{2023game}, science debate, code debugging~\cite{2024colearning}, and recommendation systems~\cite{2024macrec}. 
However, their applications in the next POI recommendation remain underexplored.

This paper proposes a Multi-Agent System for the next POI recommendation (MAS4POI). It has seven specialized agents: 1) DataAgent, 2) Manager, 3) Analyst, 4) Reflector, 5) UserAgent, 6) Searcher, and 7) Navigator.
The Manager regulates agent activities workflow and tasks allocation based on system status and resources. 
The Reflector improves recommendation quality of iterative assessment and outputs refinement.
The DataAgent organizes relevant POI data to construct accurate embeddings and trajectory visualizations. The Navigator assists route planning to generate static maps for landmark recognition.
The Analyst considers user's historical trajectory data, behavior, geographic and categorical relationships between POIs for recommendations.
The  UserAgent manages user’s preferences and the Searcher accesses external data sources and responds to user .

MAS4POI is an initial LLM-based MAS for the next POI recommendation. The system is extensible to support LLMs and external tools like \texttt{Wikipedia} and \texttt{Amap} integration. 
It can manage navigation, mitigate the cold start issues through agents collaboration with limited data for real-time Q\&A beyond POI recommendation. 
The contributions of MAS4POI are to:

\begin{itemize}
    \item propose a versatile multi-agent system to support Large Language  Models (LLMs) for the next Point-of-Interest (POI) recommendation that can  extend  seamlessly to diverse applications such as navigation and real-time question answering.
    
    \item propose seven specialized role-based agents— Manager, Reflector, Analyst, DataAgent,  UserAgent,  Searcher,  and  Navigator with each contributes to a collaborative process of task execution and iterative refinement.
    
    \item mitigate the cold start issues and validate system effectiveness on large-scale real-world datasets.
\end{itemize}

\section{Related Work}
\label{sec:related work}
\vspace{1mm}
\noindent
\highlight{Next POI Recommendation.}
Traditional POI recommendation methods rely heavily on feature engineering like collaborative and content-based filtering to extract patterns such as check-in records and ratings from historical data ~\cite{2023new, wu2024deepfeatureembeddingtabular}. 
However, they struggle with large-scale dynamic data, cold-start issues, and often overlook environmental and temporal influences on travel preferences~\cite{2022poisurvey}. 
Deep learning (DL) models like GNNs enhance recommendation quality by capturing complex spatial relationships and user interactions~\cite{2023spatio,2022getnext} requiring extensive labeled datasets, high computational costs, lack interpretability which hinder strategic planning and users trust to interact in real-time~\cite{2022poisurvey}.
Large Language Models (LLMs) have advantages in processing LBSN data textual features with embedded commonsense knowledge and broad understanding of everyday concepts ~\cite{llm4poi2024,llmmove2024,2024largellm,2023llmsurvey}. 
LLMs unify the heterogeneous data types processing to maintain contextual integrity. 
They redefine the next POI recommendation task as an intuitive question-answering mode to mitigate cold-start issues effectively~\cite{llm4poi2024}, and their complex reasoning capabilities can interpret geographical correlations and sequential transitions in user movements often overlooked by traditional models ~\cite{2024largellm}. 
Although it remains difficult to fully capture geographical contexts and mitigate aberration but they have zero-shot recommendation scenarios potentials ~\cite{llmmove2024}.  
LLMs can improve POI recommendations’ precision and clarity by  integrating  long-term  and current user preferences with spatial analysis  for personalized location-based services~\cite{2023llmsurvey}. 
MAS4POI framework is based on these strengths to specialize LLMs into distinct functional agents, collaborate with planning, discussion, decision-making, mirroring human group endeavour in problem-solving real-world interactions. 

\vspace{1mm}
\noindent
\highlight{Multi-Agent Collaboration} 
In multi-agent collaboration~\cite{2023rise}, individual agents assess the requirements and capabilities of other agents and seek engagement on cooperative actions and information sharing.
This approach can improve task efficiency, collective decision-making and address unsolved real-world issues by single agent i.e. the next POI recommendation.
Specifically, downstream agents can focus on upstream agents’ outputs when they follow certain rules for  organization~\cite{2023collaboration}. 
For example, AutoGen~\cite{2023autogen} and CAMEL~\cite{2023camel} support the strengths of individual agents to foster cooperative relationships, while AgentVerse~\cite{2023agentverse} assembles  adaptive  agent teams  dynamically based  on task  complexity.  MetaGPT ~\cite{2023metagpt} standardizes agent inputs/outputs into engineering documents and encodes them into agent prompts to structure collaboration among multiple agents  inspired  by the classical waterfall model in software development. 
But, the lack of corresponding cooperative rules, frequent interactions between multiple agents can amplify aberrations indefinitely to hinder collaboration. 
Furthermore, these multi-agent frameworks are unexplored in the next POI recommendation scenarios.

\vspace{-0.5em}
\section{Problem Statement}
\vspace{-0.5em}
\label{sec:problem_statement}
The proposed MAS4POI is to construct a multi-agent collaboration system  formalization as follows:
Consider a set of Points of  Interest  (POIs), denotes as \( P = \{p_{1}, p_{2}, \ldots, p_{M}\} \), where each POI is represents as \( \langle \text{id}, \text{cat}, g_{lat}, g_{lon} \rangle \) and \( M \) represents the number of distinct POIs. Here:
\begin{itemize}
\vspace{-0.5mm}
    \item \textit{id} refers to the unique identifier for each POI;
    \item \textit{cat} denotes the POI category (e.g., "Accommodations", "Residential"), providing semantic information;
    \item \( g_{lat}, g_{lon} \) specify the geo-coordinates (latitude and longitude).
\end{itemize}
\vspace{-0.5mm}
Each check-in record is represented as a tuple \( c_{p, t}^{u} = \langle u, p, t \rangle \in U \times P \times T \), where:
\begin{itemize}
\vspace{-0.5mm}
    \item \( u \) denotes a unique user from the user set \( U = \{u_{1}, u_{2}, \ldots, u_{N}\} \), where \( N \) represents the total number of users;
    \item \( p \) denotes a POI from the set \( P = \{p_{1}, p_{2}, \ldots, p_{M}\} \);
    \item \( t \) denotes the timestamp from the set \( T = \{t_{1}, t_{2}, \ldots, t_{Z}\} \) of each check-in record, where \( Z \) represents the number of distinct time stamps.
\end{itemize}
\vspace{-0.5mm}
A trajectory is formed as \( T_{u} = \{c_{p_{1}, t_{1}}^{u}, c_{p_{2}, t_{2}}^{u}, \ldots, c_{p_{M-1}, t_{Z-1}}^{u}, c_{p_{M}, t_{Z}}^{u}\} \), represents a sequence of POIs visited by user \( u \) at time stamp \(t \). 
For simplicity, the trajectory \( T \) denotes as \( T = \{p_{1}, p_{2}, \ldots, p_{M}\} \). 
The user’s check-in sequence is segmented into a set of consecutive trajectories
\( \{T_1^u, T_2^u, \ldots, T_k^u\} \), where each trajectory \( T_j^u = \{c_{p_{m_j}, t_{m_j}}^{u}, \ldots, c_{p_{m_j+l_j-1}, t_{m_j+l_j-1}}^{u}\} \) represents a distinct sequence of POIs visited by user
\( u \in U \) in a short time interval (e.g., 24 hours).
Given a historic trajectory \( T' = \{c_{p_1, t_1}^{u}, c_{p_2, t_2}^{u}, \ldots, c_{p_m, t_z}^{u}\} \) of user \( u \), MASPOI task is to predict the probable subsequent POI \( p_{m+1} \) to improve location-based services relevance and personalization.
\vspace{-0.5em}
\section{Method}
\begin{table}[t]
    \centering
    \vspace{-1.5em}
    \caption{Selected agents for the three MAS4POI applications. A \checkmark\ indicates a required agent, while a {$\bigcirc$} indicates an optional one.}
    \begin{tabularx}{\textwidth}{>{\centering\arraybackslash}m{1.5cm}|>{\centering\arraybackslash}m{1.5cm}>{\centering\arraybackslash}m{1.5cm}>{\centering\arraybackslash}m{1.5cm}>{\centering\arraybackslash}m{1.5cm}>{\centering\arraybackslash}m{1.8cm}>{\centering\arraybackslash}m{1.4cm}}
    \toprule
    Task & User & Analyst & Reflector & Searcher & DataAgent & Navigator \\
    \midrule
    RE & \checkmark & \checkmark & $\bigcirc$ & $\bigcirc$ & \checkmark & \\
    QA & \checkmark & $\bigcirc$ & $\bigcirc$ & \checkmark & &  \\
    NA & \checkmark &  & $\bigcirc$ & $\bigcirc$ &  & \checkmark \\
    \midrule 
    \end{tabularx}
    \label{tab:agent_selection}
    \vspace{-1.5em}
\end{table}
\vspace{-5mm}
\begin{figure}[h!]
    \centering
    \caption{The Overall Framework of MAS4POI in Next POI Recommendation Task.}
    \makebox[\textwidth][c]{%
\includegraphics[width=1.2\textwidth]{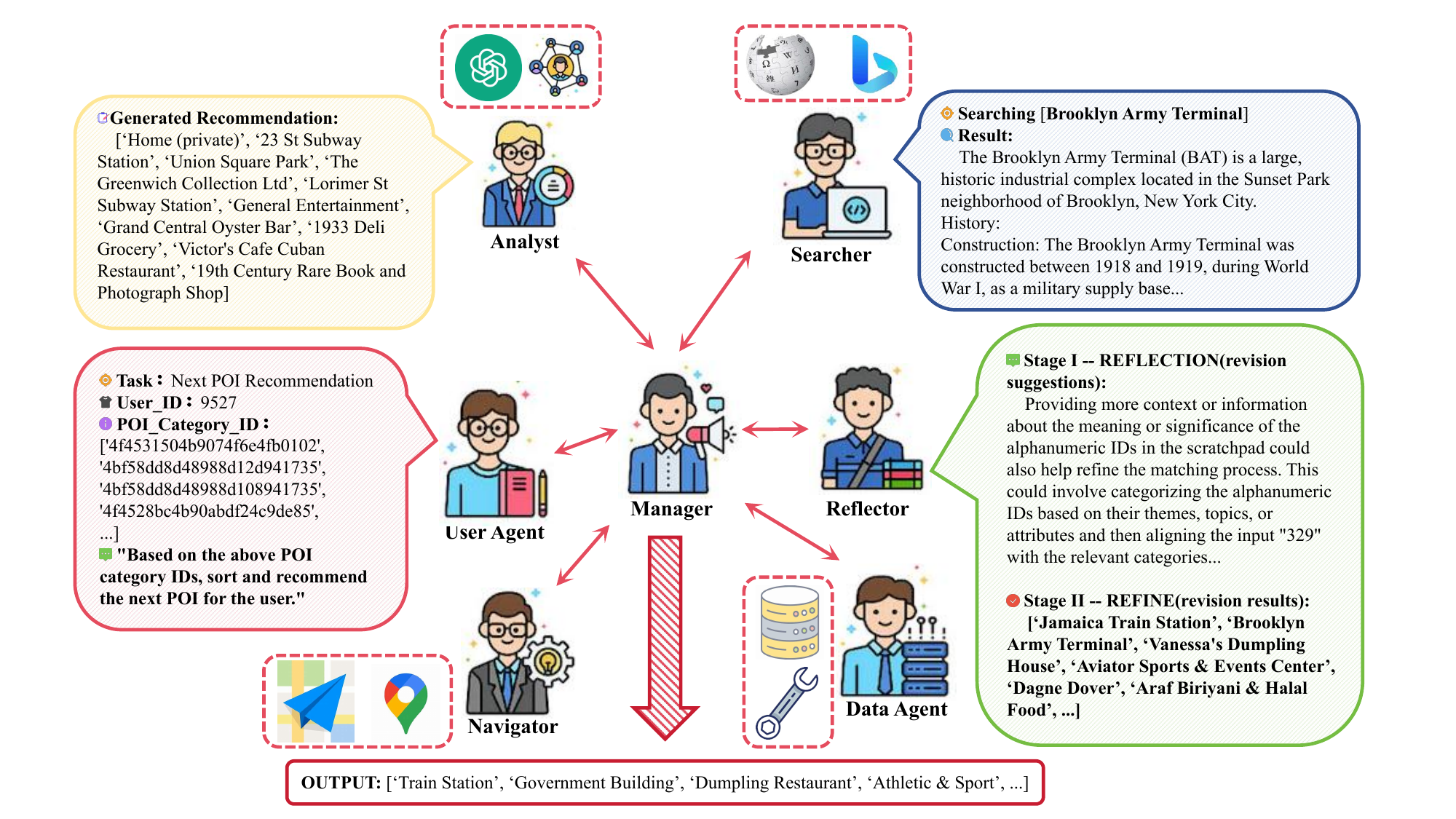}
    }
    \label{fig:framework}
    \vspace{-2.0em}
\end{figure}
\vspace{1mm}
\subsection{Framework Overview.}
A MAS4POI system overall framework is illustrated in Fig.~\ref{fig:framework}. 
It comprises of three primary applications: Next POI Recommendation, Q\&A, and Navigation.
The next POI Recommendation is the core of the workflow with Q\&A and Nav- igation as supplementary tasks with details as follows:
\vspace{-0.5em}
\subsubsection*{Next POI Recommendation:}
This process begins with the DataAgent to preprocess POI data input and constructs various check-in records based on available information. 
The Manager is a central component of the system to deliver the preprocessed data to the Analyst so that it can examine user’s past trajectory and generate an initial recommendation for the next POI. 
This recommendation is subsequently delivered to the Reflector for relevance and accuracy assessment. 
The Reflector suggests modifications if discrepancies are identified. 
Finally, the Manager conveys the refined recommendation to the UserAgent to interact with human user.
\vspace{-1.5em}
\subsubsection*{Q\&A:}
The Manager can invoke the Searcher to  manage specific queries from human in parallel so that the Searcher can access search engines to retrieve relevant information and summarize responses.
\vspace{-1.5em}
\subsubsection*{Navigation:} 
Once the user confirms the next POI, the  Manager engages the Navigator to initiate the navigation process. 
The Navigator generates a static map by the UserAgent.

An  overview  of  the  agents  selected  for  each  scenario  in MAS4POI is listed in Table~\ref{tab:agent_selection}. 
The following sections describe each agent’s  characteristics and functions in details.
\vspace{-0.5em}
\subsection{Role Agents.}
\subsubsection{Manager}
coordinates and optimizes system performance through two primary states: monitoring and operational. 
In the monitoring state, it monitors various agents’ progresses and awaits tasks completion:
\begin{equation}
\vspace{-0.5mm}
    M_{\text{monitor}}(\tau) = \prod_{a=1}^{N} \left(1 - \delta(D_a(\tau))\right)
\end{equation}
where \( \delta(D_a(\tau)) \) is an indicator function that returns 1 if the task assigned to agent \( a \) is incomplete at the given time, and 0 otherwise.
Then Manager allocates tasks based on the current system state \( S(\tau) \) and available resources \( R(\tau) \) in the operational state:
\begin{equation}
    Allocates_i(\tau) = f(S(\tau), R(\tau))
\end{equation}
Here, \( Allocates_i(\tau) \) represents the sub-tasks assigned to agent \( a \).
The Manager effectively coordinates the activities of other agents within the system by controlling these assignments.
\vspace{-5mm}
\begin{algorithm}
\caption{\textit{Reflector} with in MAS4POI, all details have been discussed in section ~\ref{method:reflector}}
\label{al:reflector}
\begin{algorithmic}[1]
\Require Input $x$, Manager $\mathcal{M_{a}}$, Reflector $\mathcal{R_{a}}$, prompts $\{p_{m}, p_{th}, p_{re}\}$, number of iterations $\mathcal{N}$, Ending condition $\text{end}(\cdot)$
\Ensure Corrected Next POI Recommendation $\hat{y}$
\State Generate initial output $y_0 = \mathcal{M_{a}}(p_{m} \parallel x)$ \Comment{Initialization (Eqn.~\ref{eq:initial})}
\For {$i \leftarrow 0$ \textbf{to} $\mathcal{N}-1$ }
    \State $Ref_{i} = \mathcal{R_{a}}(p_{th} \parallel x \parallel y_i)$ \Comment{Reflection (Eqn.~\ref{eq:reflection})}
    \If {$\text{end}(Ref_{i}, i)$} \Comment{Ending Condition}
        \State \textbf{break}
    \Else
        \State $y_{i+1} = \mathcal{R_{a}}(p_{re} \parallel x \parallel y_0 \parallel Ref_{0} \parallel \ldots \parallel y_i \parallel Ref_{i})$ \Comment{Refine (Eqn.~\ref{eq:iterating})}
    \EndIf
\EndFor
\State \Return $y_i$
\end{algorithmic}
\end{algorithm}
\vspace{-8mm}
\subsubsection{Reflector}
\label{method:reflector}
are crucial processes for agent’s self-assessment and iterative improvement. 
It focuses on Reflection and Refinement of the Manager’s outputs to identify improvement areas and optimize the overall process.
Given input \( x \), the \textit{Manager} \( \mathcal{M_{a}} \) generates initial output \( y_0 \) based on the prompt \( p_{m} \).
\vspace{-0.5em}
\begin{equation}
\label{eq:initial}
\vspace{-0.5em}
    y_0 = \mathcal{M_{a}}(p_{m} \parallel x)
\end{equation}
Reflector proposes reflections \( Ref_{i} \) in the REFLECTION process based on the reflection prompt \( p_{th} \).
\vspace{-0.5em}
\begin{equation}
\label{eq:reflection}
    Ref_{i} = \mathcal{R_{a}}(p_{th} \parallel x \parallel y_i)
\end{equation}
After REFLECTION, the \textit{Reflector} refines the prediction strategy and candidate selection based on the reflection prompt \( p_{re} \) to generate an accurate subsequent output \( y_{i+1} \) in \textbf{REFINE}.
\vspace{-0.5em}
\begin{equation}
\label{eq:refine}
    \vspace{-0.5em}
    y_{i+1} = \mathcal{R_{a}}(p_{re} \parallel x \parallel y_t \parallel Ref_{i})
\end{equation}
Finally, \textit{Reflector} alternates between REFLECTION and REFINE until an ending condition (\( \text{end}(Ref_{i}, i) \)) is met. 
This condition, defined by \( \text{end}(Ref_{i}, i) \), can either be a specified iteration limit or a stopping indicator triggered by the reflection results.
\vspace{-1mm}
\begin{equation}
\label{eq:iterating}
    y_{i+1} = \mathcal{R_{a}}(p_{re} \parallel x \parallel y_0 \parallel Ref_{0} \parallel \ldots \parallel y_i \parallel Ref_{i})
\end{equation}

To ensure the \textit{Reflector} learns from previous iterations and avoids repeating mistakes, the history of all prior feedback and outputs are retained, appended to the prompt and stored in short-term memory during each iteration. 
The last refined output \( \hat{y} \) is used as the final recommendation. The overall process is listed in Algorithm~\ref{al:reflector}.
\vspace{-1.5em}
\subsubsection{DataAgent}
\label{md:data}
preprocess and structure check-in data integral for precise POI recommendations. 
For each user \( u \), the raw check-in 
records sequence \( C_u = \{c_{p_1, t_1}^{u}, c_{p_2, t_2}^{u}, \ldots, c_{p_n, t_n}^{u}\} \) is processed.
The DataAgent filters out POIs and users with fewer than 10 visit records by equation~\ref{eq:filter}, leading to the filtered sequence:
\vspace{-0.5mm}
\begin{equation}
\label{eq:filter}
    C_u^{filtered} = \{c_{p_i, t_i}^{u} \mid \text{count}(c_{p_i, t_i}^{u}) \geq 10\}
\end{equation}
DataAgent aggregates filtered data into structured POI information \( P \) subsequently, including geo-coordinates and categories, and segments the user’s check-ins into trajectories \( \{T_1^u, T_2^u, \ldots, T_k^u\} \), where each \( T_j^u \) comprises of a list of POIs visited within a 24-hour interval. 
The comprehensive dataset is then formed as:
\begin{equation}
    \text{Check-in Data} = \bigotimes_{u \in U} \left(\{C_u^{filtered}\}, \{P_m^{filtered}\}\right)
\end{equation}
to improve POI recommendations reliability and accuracy.
\vspace{-1.5em}
\subsubsection{Navigator}
improves user experience by providing precise route planning and visual route maps through \texttt{Amap} APIs integration. 
This process begins with geocoding user-provided addresses into geographic coordinates. 
After obtaining these coordinates, Navigator calculates the optimal route using the selected mode of transportation with the Haversine Formula for precise distance measurement:
\vspace{-1mm}
\begin{equation}
\label{eq:haversine}
    \Delta d = R \cdot \Delta \sigma
    \vspace{-1mm}
\end{equation}
where \(\Delta \sigma\) represents the angular distance in radians, determined by:
\begin{equation}
    \Delta \sigma = 2 \cdot \arcsin \left(\sqrt{\sin^2 \left(\frac{\Delta \phi}{2}\right) + \cos(\phi_1) \cdot \cos(\phi_2) \cdot \sin^2 \left(\frac{\Delta \lambda}{2}\right)}\right)
\end{equation}
Here, \(\Delta \phi\) and \(\Delta \lambda\) are the differences in latitude and longitude, and \(R\) is the Earth's radius. 
Navigator also generates static maps to visually represent routes for landmark recognition and effective navigation.
\vspace{-1mm}
\subsubsection{Analyst}
\vspace{-1.5em}
is based on the user’s past visit records
\( \mathcal{H}_u \) (consist of long-term and recent records) and candidate set \( \mathcal{C}_u \), which are defined as follows:
\begin{equation}
\label{eq:history_and_candidates}
\mathcal{H}_u = \{(p_i, \text{cat}(p_i), t_i) \mid i = 1, 2, \ldots, H\}, \quad
\mathcal{C}_u = \left\{ \left(p_j, \Delta d(p_j, p_{\text{last}}), \text{cat}(p_j)\right) \right\}_{j=1}^{M'}
\end{equation}
where \( H \) indicates total historic records and \( p_i \) denotes a visited POI defined in~\ref{sec:problem_statement}, \( \text{cat}(p_i) \) is its category, and \( t_i \) is the visit timestamp. \( \Delta d(p_j, p_{\text{last}}) \) represents the distance 
(calculated through~\ref{eq:haversine}) between a candidate POI \( p_j \) and the last visited POI \( p_{\text{last}} \), and \( \text{cat}(p_j) \) represents the candidate POI's category.
The \textit{Analyst} combines \( \mathcal{H}_u \) and \( \mathcal{C}_u \) into a prompt \( p_{an} \) to generate the list of initial recommended POIs with explanation.
\begin{equation}
y_{\mathcal{A_{a}}} = \mathcal{A_{a}} (p_{an} \parallel \mathcal{H}_u, \parallel \mathcal{C}_{u})
\end{equation}

\vspace{-2.0em}
\subsubsection{UserAgent}
interacts with the user to collect requirements, store user accounts, and retrieve user historical records to generate initial POI recommendations.
\vspace{-1.5mm}
\subsubsection{Searcher} 
The \textit{Searcher} processes the user's query \( q \) using a set of tools \( \mathcal{T}_S \) with the Searcher prompt \( p_{se} \) to generate the final response \(y_{\mathcal{S_{a}}} \):
\begin{equation}
y_{\mathcal{S_{a}}} = \mathcal{S_{a}} \left(q \parallel \mathcal{T}_S \parallel p_{se}\right)
\end{equation}
where $s$ indicates the total number of tools and \( \mathcal{T}_S \) is defined as:
\begin{equation}
\vspace{-0.5mm}
    \mathcal{T}_S = \{ \text{tool}_{1}, \text{tool}_{2}, \ldots, \text{tool}_{s} \}
    \vspace{-3mm}
\end{equation}
\vspace{-5mm}
\section{Experiments}
\subsection{Experiment Setup}
\vspace{-0.5em}
\noindent
\highlight{Dataset.}
There are two real-world datasets derived from location-based social media platforms for the experiments: NYC(Foursquare-NYC) and TKY (Foursquare-TKY) are. NYC contains  103,941 check-ins (check in records?)collected from 1,048 users across 4,981 POIs. TKY includes 405,000 check-ins from 2,282 users across 7,833 POIs. Both datasets are collected over approximately from April 2012 to February 2013 as listed in Table~\ref{tab:dataset information}.
\begin{table}[t]
    \small
    \centering
    \renewcommand{\arraystretch}{1.}
    \caption {Dataset information.}
    \label{tab:dataset information}
    \resizebox{0.6\columnwidth}{!}{
        \begin{tabular}{ c | c | c | c | c | c }
    \toprule[1.5pt]
    Name &  Users &  POIs  &  Categories & Check-ins & Trajectory \\
    \midrule[1pt]
    NYC  & 1,048   & 4,981  & 318  & 103,941  & 14,130 \\ 
    TKY  & 2,282   & 7,833  & 290  & 405,000  & 65,499 \\ 
    \bottomrule[1.5pt]
        \end{tabular}
    }
    \vspace{-4mm}
\end{table}

\vspace{1mm}
\noindent
\highlight{Baseline Methods.}
The experiments used six distinct LLMs, including GLM-3-Turbo ~\cite{chatglm2024}, GPT-3.5-Turbo~\cite{openai2023}, MoonShot-v1, QWEN-turbo~\cite{qwen2023}, Claude-3.5, and Gemini-Pro~\cite{gemini2023}. 
They are integrated MAS4POI for evaluation. 
\vspace{1mm}
\noindent
\highlight{Evaluation Metrics.}
MAS4POI is evaluated by two standard metrics: Acc$@k$ and Mean Reciprocal Rank (MRR) as described in ~\cite{llmmove2024,llm4poi2024}.
Acc$@k$ analyses the correct POI is within the top-k recommendations as an unordered list. 
$MRR$ considers the correct POI position is within the ordered list. For a dataset with M sample trajectories, the metrics are defined as follows:
\[
\begin{minipage}{0.45\textwidth}
\centering
\vspace{-4mm}
\[
\text{Acc@k} = \frac{1}{M} \sum_{i=1}^{M} 1(r_i < k)
\]
\end{minipage}
\hspace{0.05\textwidth}
\begin{minipage}{0.45\textwidth}
\centering
\vspace{-3mm}
\[
\text{MRR} = \frac{1}{M} \sum_{i=1}^{M} \frac{1}{rank_i}
\]
\end{minipage}
\]
Where $rank_{i}$ denotes the correct next POI position in the recommended list.
Generally, higher values indicate the recommendation system’s improvements.
\vspace{1mm}
\noindent
\highlight{Implementation Details} 
There are five experiments conducted and averaged the results for each dataset as listed in Table~\ref{tab:performance_comparison}. The temperature of all LLMs is set at 0 for comparison. The datasets preprocessing  is managed by DataAgent, and are partitioned into training, validation, and test sets with a ratio of 8:1:1. Notably, only the final check-in record of each trajectory is assessed in the validation and test sets.
\vspace{-1.5em}
\subsection{Main Results}
\vspace{-0.5em}
MAS4POI integrates six LLMs across three datasets experiments results are listed in Table ~\ref{tab:performance_comparison}. 
For NYC dataset, it is noted that the Claude model achieved the best $Acc@1$ and $MRR$ scores 0.8172 and 0.8251 respectively, which are  1.6\%  and  0.55\%  higher than the second-best  Gemini  model accordingly.
It is noted that Claude’s $Acc@5$ and $Acc@10$ are  near  to  Gemini’s scores 0.8325 and 0.8470.
While Gemini surpasses their scores 0.8333 and 0.8517, indicating that Claude are slightly advantaged than Claude’s $Acc@5$ and $Acc@10$.
For TKY dataset, it is noted that the  QWEN model achieved the best $Acc@1$ and $MRR$ scores 0.8831 and 0.9279 respectively, which are higher than other models.
QWEN outperformed Gemini in MRR by 4.0\%, indicating that its relevance and accuracy. 
It is also noted that the MoonShot model achieved average $Acc@1$ and MRR scores 0.7304 and 0.7520 on the NYC dataset respectively.
For TKY dataset, the MoonShot model has achieved $Acc@1$ and $MRR$ scores 0.6533 and 0.6742 respectively, indicating that MoonShot has deficiencies in $Acc@k$ and $MRR$.
It is also noted that the Claude model performed satisfactory at the NYC dataset, while the QWEN model performed satisfactory at the TKY dataset.
Hence, it is considered that Claude and QWEN are suitable LLMs for MAS4POI. It is also noted that the GPT model is widely used in various fields ~\cite{ChatGPT2023} but the performance between GPT and QWEN at the TKY dataset is insignificant at $Acc@k$ (k=1, 5, 10) with 2.11\%, 1.31\%, and 1.0\% differences. 
Hence, GPT remains as the primary LLM for MAS4POI.
Therefore, we recommend Claude and QWEN as the main LLM choices for MAS4POI. 

\begin{table*}[t]
\centering
\caption{MAS4POI Performance Comparison of Abbreviated LLMs Across NYC and TKY Datasets Using $Acc@k$ (k=1, 5, 10) and $MRR$ Metrics.}
\renewcommand{\arraystretch}{1.5} 
\setlength{\tabcolsep}{10pt} 
\resizebox{1.0\textwidth}{!}{
    \begin{tabular}{lccccccccc}
    \hline
     & \multicolumn{4}{c}{NYC} & \multicolumn{4}{c}{TKY}\\
    \cline{2-5} \cline{7-10}
    LLMs & $Acc@1$ & $Acc@5$ & $Acc@10$ & $MRR$ & & $Acc@1$ & $Acc@5$ & $Acc@10$ & $MRR$ \\
    \hline
    GLM   & 0.7572 & 0.7640 & 0.7871 & 0.7508 & & 0.8317 & 0.8693 & 0.8892 & 0.8651 \\
    GPT  & 0.7680 & 0.7940 & 0.8070 & 0.7691 &  & 0.8620 & 0.8936 & 0.9033 & 0.8748 \\
    MoonShot   & 0.7304 & 0.7537 & 0.7633 & 0.7520 & & 0.6533 & 0.7233 & 0.7302 & 0.6742  \\
    Claude  & \textbf{0.8172} & 0.8325 & 0.8470 & \textbf{0.8251} & & 0.8256 & 0.8970 & 0.9051 & 0.8510  \\
    Gemini  & 0.8012 & \textbf{0.8333} & \textbf{0.8517} & 0.8196 & & 0.8721 & \textbf{0.9164} & \textbf{0.9217} & 0.8879 \\
    QWEN   & 0.7821 & 0.8140 & 0.8468 & 0.8250 & & \textbf{0.8831} & 0.9067 & 0.9133 & \textbf{0.9279} \\
    \hline
    \end{tabular}
    }   
\label{tab:performance_comparison}
\vspace{-2.0em}
\end{table*}
\vspace{-4mm}
\begin{table*}[t]
\centering
\caption{User cold-start analysis on the NYC and TKY datasets, GPT within MAS4POI.}
\renewcommand{\arraystretch}{1.5} 
\setlength{\tabcolsep}{10pt} 
\resizebox{1.0\textwidth}{!}{
\begin{tabular}{lccccccccc}
    \hline
     & \multicolumn{4}{c}{NYC} & & \multicolumn{4}{c}{TKY} \\
    \cline{2-5} \cline{7-10}
    User Group & $Acc@1$ & $Acc@5$ & $Acc@10$ & $MRR$ & & $Acc@1$ & $Acc@5$ & $Acc@10$ & $MRR$ \\
    \hline
    Inactive   & 0.7817 & 0.8150 & 0.8267 & 0.7943 & & 0.8100 & 0.8533 & 0.8633 & 0.8887 \\
    Normal  & 0.7733 & 0.8132 & 0.8333 & 0.7852 &  & 0.8367 & 0.8733 & 0.8900 & 0.8524 \\
    Very\_active  & 0.8167 & 0.8500 & 0.8533 & 0.8251 & & 0.8233 & 0.8700 & 0.8733 & 0.8925  \\
    \hline
\end{tabular}
}
\label{tab:cold_start}
\vspace{-1.5em}
\end{table*}
\subsection{User Cold Start Analysis.}
\vspace{-1mm}
Users are categorized into three groups based on the trajectories number: \textit{inactive} (bottom 30 in trajectory count), \textit{normal}, and \textit{very\_active} (top 30 in trajectory count) in the training set to evaluate MAS4POI in mitigating the cold start issues. The experiment results are listed in Table~\ref{tab:cold_start}.
It showed that the inactive users for $Acc@k$ (k=1,5,10) and $MRR$ is slightly lower than the \textit{very\_active}  users at the NYC dataset. 
For example, the $Acc@10$ and $MRR$ difference between the inactive and the \textit{very\_active} users at the NYC dataset is only 2.66\% and 3.08\% respectively. 
For TKY dataset, the gap is also limited: the inactive users at $Acc@1$ scores 0.8100, compared to the \textit{very\_active} users scores 0.8233. 
For $Acc@10$ and $MRR$, the inactive users is near or slightly lower than the \textit{very\_active} users, which underscored MAS4POI to mitigate the cold start issues for the inactive users effectively. 
This is attributed to the collaboration between multiple agents. 
The overlap in POI trajectories among different users allows the Analyst to support the structured check-in records constructed by the DataAgent to identify suitable candidate POIs, even if an inactive user’s own trajectory is limited i.e. users often visit CaffeeShop or PublicTransitStations at weekday mornings. 
Additionally, the Reflector enhances system’s accuracy by correcting and refining the Manager’s final output. 
The Reflector experiments are described in ~\ref{sec:ablation}.

\begin{figure}[t]
    \makebox[\textwidth][c]{%
        \includegraphics[width=1.0\textwidth]{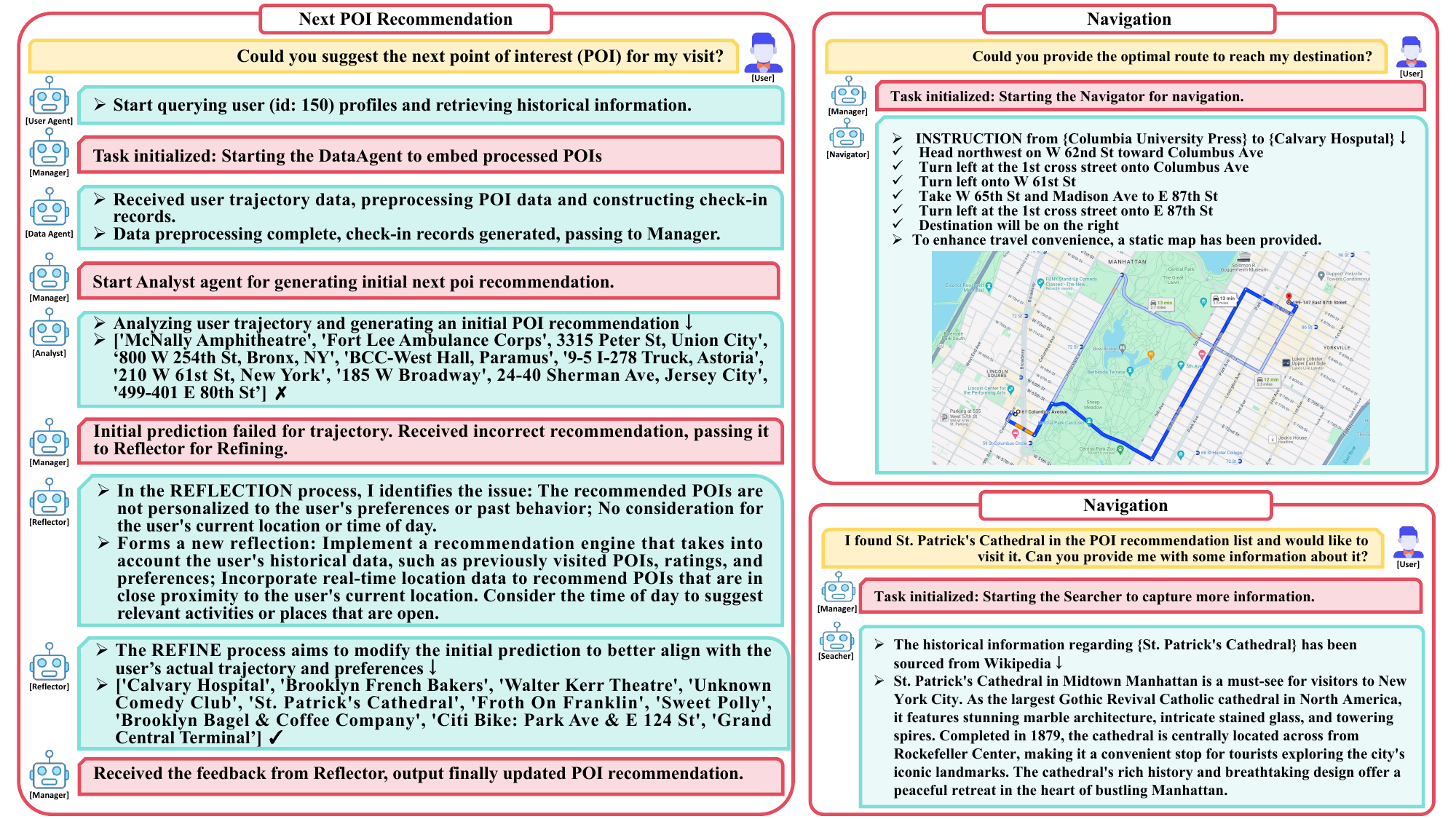}
    }
    \caption{MAS4POI Workflow with Key Elements Highlighted (Red: incorrect POI recommendations initially generated by \textit{Analyst}, Blue: Refined POI recommendation output, Orange: REFLECTION process carried out by \textit{Reflector}, Green: API Results, Purple: User Requests)}
    \label{fig:case}
    \vspace{-4mm}
\end{figure}
\vspace{-1mm}
\subsection{Case Study}
\label{sec:case}
A MAS4POI’s workflow using the NYC dataset is illustrated in Fig.~\ref{fig:case}. 
It showed that  the Manager oversaw the Analyst to generate initial recommendations from user’s requirements on the next POI recommendation task, but these recommendations often have geographical mismatches.
For example 1) 24\~40 Sherman Ave is located at Jersey City, does not meet the user’s requirement for POIs current location (New York), 2) time factors are disregarded such as  BCC West Hall was not near to the public at specific times. Hence, the Reflector corrects these recommendations by focusing on the user’s real-time location to prioritize nearby POIs. 
For example, the current POI to CalvaryHospital is only an 11-minute drive. Additionally, the user$_{id = 150}$ frequently visits places about performing arts such as PerformingArtsVenue and Theater, resulting WalterKerrTheatre is a personalized recommendation based on this context since user$_{id = 150}$ complete historical trajectories are provided by the system’s open-source code. 
For Q\&A, MAS4POI supports user queries through the Searcher. 
For example, when the user expresses interest in St. Patrick's Cathedral the Searcher retrieves detailed information and summarizes historical data from a reliable source like \texttt{Wikipedia} and \texttt{Bing}.
Additionally, the Navigator generates an optimal route with clear instructions and a static map for user’s decision-making.
\vspace{-2mm}
\subsection{Ablation Study}
\vspace{-1mm}

\begin{figure}[H]
\centering
\vspace{-1.5mm}
\caption{\textbf{Upper}: This table shows the improvement in $Acc@k$ and MRR with different states based on GPT-3.5-Turbo, when state equals to $y_{0}$, it indicates no use of \textit{Reflector}. \textbf{Below}: This shows performance improvements with iterations.} 
\begin{minipage}[H]{1.0\textwidth}
    \centering
    \renewcommand{\arraystretch}{1.5} 
    \setlength{\tabcolsep}{10pt} 
    \resizebox{\textwidth}{!}{
        \begin{tabular}{lccccccccc}
        \hline
         & \multicolumn{4}{c}{NYC} & & \multicolumn{4}{c}{TKY} \\
        \cline{2-5} \cline{7-10}
        States & $Acc@1$ & $Acc@5$ & $Acc@10$ & $MRR$ & & $Acc@1$ & $Acc@5$ & $Acc@10$ & $MRR$ \\
        \hline
        $y_{0}$   & 0.7166 & 0.7566 & 0.7833 & 0.7363 & & 0.7933 & 0.8433 & 0.8567 & 0.8161\\
        $y_{1}$   & 0.7466 & 0.7767 & 0.7933 & 0.7588 & & 0.8476 & 0.8798 & 0.8930 & 0.8699 \\
        $y_{2}$   & 0.7601 & 0.7882 & 0.8012 & 0.7643 & & 0.8613 & 0.8872 & 0.9000 & 0.8720 \\
        $y_{3}$   & 0.7680 & 0.7940 & 0.8070 & 0.7691 &  & 0.8620 & 0.8936 & 0.9033 & 0.8748 \\
        \hline
        \end{tabular}
    }
\end{minipage}%
\hspace{0.1\textwidth} 
\begin{minipage}[H]{1.0\textwidth}
    \centering
    \includegraphics[width=\linewidth]{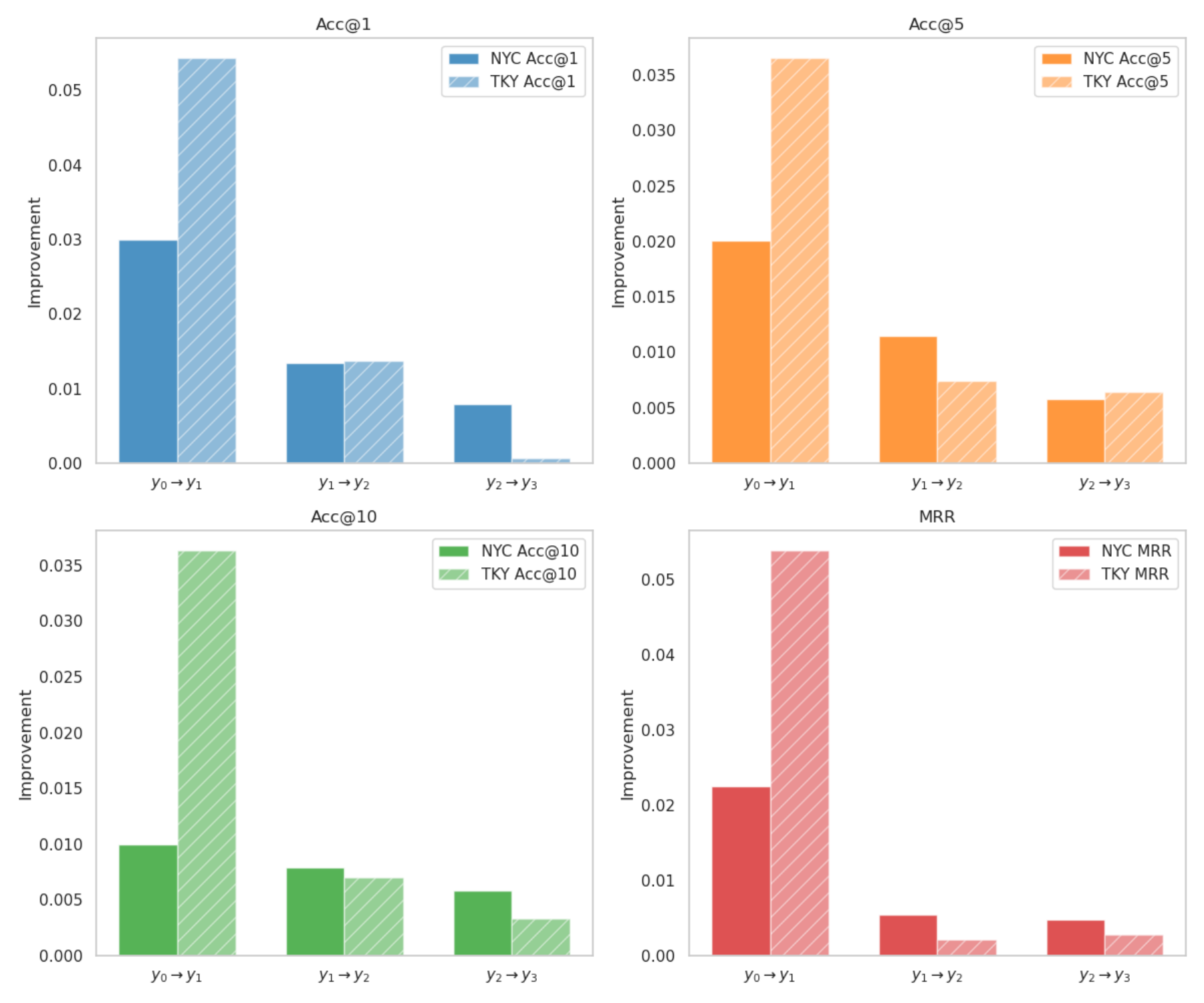}
\end{minipage}
\label{fig:ablation_combined}
\end{figure}

\label{sec:ablation}
This section examines the \textbf{REFLECTION} and \textbf{REFINE} stage iterations of the Reflector in MAS4POI’s performance.
The results are illustrated in Fig.~\ref{fig:ablation_combined}.
It showed that MAS4POI improves performance across different datasets as the stages number increase. 
For example, at the TKY dataset, transitioning from state $y_{0}$  (without Reflector) to $y_{1}$  resulted 5.43\%, 3.65\%, 3.63\%, and 5.38\% at $Acc@k$ (k=1, k=5, k=10) and $MRR$ respectively. 
For the NYC dataset, the performance results increase from the initial 0.7166, 0.7566, 0.7833, and 0.7363 to 0.7680, 0.7940, 0.8070, and 0.7691  after three iterationsaccordingly. However, it is noted that the improvement gradually diminish at the MRR metric in particular and time cost increase as the Reflector’s REFLECTION and REFINE iterations number increase.

\vspace{-5mm}

\section{Conclusion} 
\vspace{-3mm}
In this research, we proposed a Multi-Agent System for next POI recommendation (MAS4POI), comprising seven role agents: \textit{DataAgent}, \textit{Manager}, \textit{Analyst}, \textit{Reflector}, \textit{UserAgent}, \textit{Searcher}, and \textit{Navigator}. 
Each agent plays a specific role: \textit{Manager} optimizes coordination, \textit{Reflector} enhances system accuracy through iterative self-assessment, \textit{DataAgent} preprocesses and organizes structured POI data, \textit{Navigator} handles route planning, \textit{Analyst} generates recommendations from user historical check-in records, \textit{UserAgent} manages user profiles and interactions, and \textit{Searcher} processes specific queries. 
We deployed six different LLMs within MAS4POI and evaluated them using Acc$@k$ ($k$=1, 5, 10) and MRR across three datasets. 
Our results demonstrate the effectiveness of MAS4POI. While LLMs introduce challenges such as hallucination and repetition, which can affect user experience, we believe these issues will mitigated as LLMs evolve. 
Finally, we have made our code publicly available to support future research and hope MAS4POI will contribute to the advancement of next POI recommendation.

\vspace{-2mm}
\section{Acknowledgment}
\vspace{-2mm}
\label{sec:acknowledgment}
The authors thank for Beijing Normal University-Hong Kong Baptist University United International College and the IRADS lab for the provision for computer facility for the conduct of this research.

\end{document}